\documentclass[aps,twocolumn,superscriptaddress,showpacs,groupedaddress,reprint]{revtex4}  % for review and submission

\usepackage{graphicx}  % needed for figures
\usepackage{dcolumn}   % needed for some tables
\usepackage{bm}        % for math
\usepackage{amssymb}   % for math

\usepackage{xspace}
\usepackage{amsmath}
\usepackage{abbrevs}
\usepackage[]{units}	%Auto-formats units
\usepackage{color}

\usepackage{float} 
\usepackage{wrapfig} 
\usepackage{multirow} % Required for multirows
\usepackage{rotating} %rotating tables

\begin{document}

\title{A fiber-integrated quantum memory for telecom light}

\author{K.~A.~G. Bonsma-Fisher}
\affiliation{National Research Council of Canada, 100 Sussex Drive, Ottawa, Ontario K1A 0R6, Canada}

\author{C. Hnatovsky}
\affiliation{National Research Council of Canada, 100 Sussex Drive, Ottawa, Ontario K1A 0R6, Canada}

\author{D. Grobnic}
\affiliation{National Research Council of Canada, 100 Sussex Drive, Ottawa, Ontario K1A 0R6, Canada}

\author{S.~J. Mihailov}
\affiliation{National Research Council of Canada, 100 Sussex Drive, Ottawa, Ontario K1A 0R6, Canada}

\author{P.~J. Bustard}
\email[]{philip.bustard@nrc-cnrc.gc.ca}
\homepage{http://quantumtechnology.ca}
\affiliation{National Research Council of Canada, 100 Sussex Drive, Ottawa, Ontario K1A 0R6, Canada}

\author{D.~G. England}
\email[]{duncan.england@nrc-cnrc.gc.ca}
\affiliation{National Research Council of Canada, 100 Sussex Drive, Ottawa, Ontario K1A 0R6, Canada}

\author{B.~J. Sussman}
\affiliation{National Research Council of Canada, 100 Sussex Drive, Ottawa, Ontario K1A 0R6, Canada}
\affiliation{Department of Physics, University of Ottawa, Advanced Research Complex, 25 Templeton Street, Ottawa, Ontario K1N 6N5, Canada}

\date{\today}

\begin{abstract}
\noindent We demonstrate the storage and on-demand retrieval of single-photon-level telecom pulses in a fiber cavity.
The cavity is formed by fiber Bragg gratings at either end of a single-mode fiber. 
Photons are mapped into, and out of, the cavity using quantum frequency conversion driven by intense control pulses. 
In a first, spliced-fiber, cavity we demonstrate storage up to 0.55\,$\mu$s (11 cavity round trips), with $11.3 \pm 0.1$\% total memory efficiency, and a signal-to-noise ratio of $12.8$ after 1 round trip. 
In a second, monolithic cavity, we increase this lifetime to 1.75\,$\mu$s (35 round trips) with a memory efficiency of $12.7 \pm 0.2\%$ (SNR of $7.0 \pm 0.2$) after 1 round trip.
Fiber-based cavities for quantum storage at telecom wavelengths offer a promising route to synchronizing spontaneous photon generation events and building scalable quantum networks. 
\end{abstract}


\maketitle

\section{Introduction}
Quantum memories are vital components for a number of emerging quantum technologies including quantum communication~\cite{Duan2001} and quantum computing~\cite{KLM}. 
A quantum memory faithfully stores a photonic state and releases it on demand. This capability can be used to synchronize spontaneous photon generation, allowing the build-up of large photon-number entangled states~\cite{Nunn2013}, a key requirement for scalable linear optical quantum computing~\cite{Varnava2008}. 
Quantum memories which operate at telecom wavelengths are of particular importance, since they can be incorporated into fiber networks and enable widespread quantum networking~\cite{Kimble2008}.
Telecom storage has been realized in a number of physical systems including rare-earth doped crystals~\cite{Dajczgewand2014, Liu2022}, erbium-doped optical fiber~\cite{Saglamyurek2015}, 
hot atomic vapor~\cite{Thomas2023}, and opto-mechanical resonators~\cite{Wallucks2020}. 
The typical strategy for quantum storage is to map a flying photon to a stationary, material excitation. 
In this way, the long decoherence times associated with well-isolated atomic transitions, measured in hours in the state-of-the-art~\cite{Zhong2015}, can be leveraged.
Storage times of more than 1\,ms have been demonstrated with telecom wavelengths~\cite{Wallucks2020}.
Another key metric for quantum memories is the storage bandwidth. 
Memory bandwidths exceeding 1\,THz for near-infrared, and 1\,GHz for telecom, wavelengths have been demonstrated using the Raman~\cite{England2015} and ORCA~\cite{Thomas2023}, respectively. 
These two metrics together give the time-bandwidth product, which, operationally, is how many time-bins a photon is stored for. 
For telecom storage, time-bin products in excess of $10^4$ have been realized~\cite{Wei2022}.
While long-lived memories will form the backbone of fiber-based long-distance quantum networks,  shorter-lived memories with high time-bandwidth products will be essential for local networking tasks~\cite{Heshami2016}, such as the multiplexing of spontaneous photon sources.

A second strategy for photon storage is to trap light in a low-loss cavity~\cite{Pittman2002}.
This has been realized in two distinct methods: free-space cavities where write and read operations can be controlled efficiently with electro-optic switches~\cite{Kaneda2017, Kaneda2019}; and fiber-based cavities where operations are performed optically using intense control pulses~\cite{Bustard2022}.
Free-space cavities have already been integrated with photon generation and scaling advantages for entangled photon states have been observed~\cite{Meyer-Scott2022}, but may be ultimately limited by loss from free-space optics. 
Fiber-based cavities may achieve lower loss levels, especially in the telecom bands where they have excellent transmission efficiency. The main hurdles are then the efficiency of write and read operations, and the mitigation of any additive noise processes.

%%% FIG %%%
\begin{figure}[ht]
\center{\includegraphics[width=1.\columnwidth]{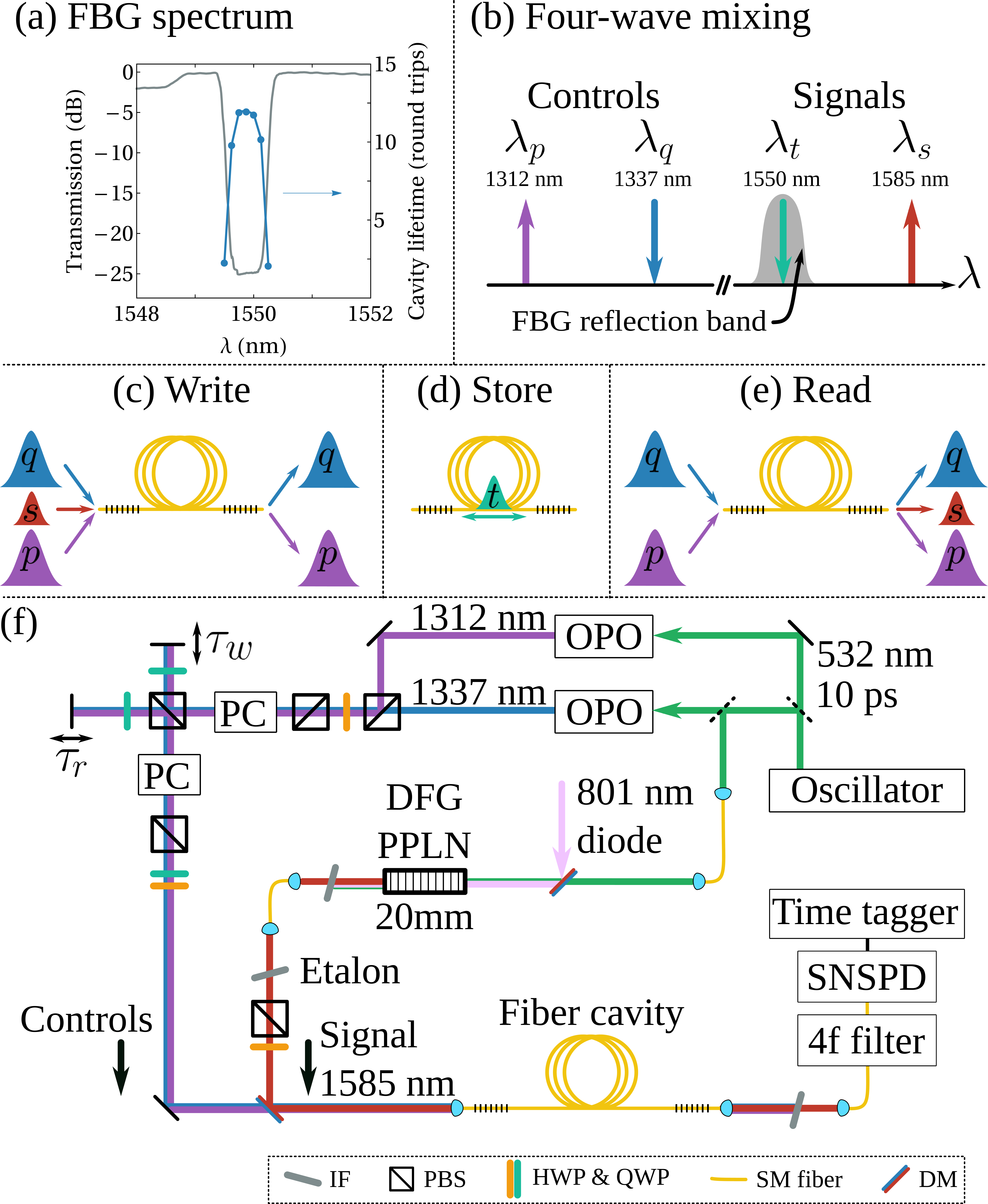}}
\caption{ (a) The transmission spectrum of the fiber Bragg gratings (FBG) (left axis) in the spliced-fiber cavity plotted along with the cavity lifetime ($1/e$ ringdown constant) as a function of wavelength (right axis). 
(b) Signal field ($s$) is frequency-converted in the fiber via BSFWM mediated by two control fields ($p$ and $q$), facilitating the read and write operations.
The wavelength of the converted light ($\lambda_t$) is in the FBG reflection band.
(c-e) Memory operation. 
A signal photon $s$ is written into the memory by the frequency conversion process (c),
where the converted pulse is trapped in the fiber cavity, reflecting off the FBGs (d).
(e) A read pulse (comprised of fields $p$ and $q$) arrives at a later time driving the reverse frequency conversion process, mapping the stored photon back to the original signal mode.  
(f) Experimental setup, see main text for details.
Optical parametric oscillator (OPO); Pockels cell (PC); difference frequency generation (DFG); periodically-poled lithium niobate (PPLN); superconducting-nanowire single photon detector (SNSPD); interference filter (IF); polarizing beamsplitter (PBS); half- and quarter-wave plate (HWP and QWP); single-mode (SM) fiber; dichroic mirror (DM).
}\label{fig:setup}
\end{figure}

In this work we develop a fiber-based quantum memory for telecom light with a 34\,GHz bandwidth for application to photon source multiplexing.
We store single-photon-level telecom pulses in a fiber cavity comprised of two fiber Bragg gratings (FBGs) written at opposite ends of a single-mode fiber. 
The operation of the memory is controlled optically using Bragg scattering four-wave mixing (BSFWM) driven by intense control pulses to convert an L-band signal photon to, and from, the storage wavelength (1550\,nm).
In a first demonstration, we form a fiber cavity by splicing together two commercially-available FBGs. 
We retrieve a signal photon from the memory for up to 11 cavity round trips of storage, corresponding to 0.55\,$\mu$s of storage time. 
The retrieved signal shows high spectral and temporal overlap with the input, and retains a signal-to-noise ratio $\geq 1$.
To assess the memory's ability to time-multiplex photons, we interfere a photon retrieved from the memory with an externally delayed photon, showing high interference visibility.
We find that the storage lifetime of this cavity is ultimately limited by the splice and so, in a second experiment, we fabricate a home-built monolithic cavity by inscribing FBGs at either end of a single-mode fiber. 
The monolithic cavity shows significantly longer storage times: up to 35 round trips, or 1.75\,$\mu$s of storage time. 
We also demonstrate that the monolithic cavity can be mechanically stretched into resonance with the primary laser repetition rate, which is vital for multiplexing single photon sources. 
The storage in our monolithic cavity is ultimately limited by dispersion from the FBGs and unwanted polarization rotation of the stored signal.

\section{ Spliced-fiber cavity } 
Fibre Bragg gratings~\cite{Kashyap2010} are formed by increasing the index of refraction in a fiber core ($n_\text{core} + \Delta n$) in a periodic structure along the propagation axis of guided light. 
The period of the grating ($\Lambda$) causes forward-travelling light at the Bragg wavelength $\lambda_B=2 n_\text{eff} \Lambda$ to reflect into the backward-travelling mode, where $n_\text{eff}$ is the effective index of refraction along the length of the FBG. 
The reflectivity, bandwidth, and dispersive properties of the FBG can be further customized by using non-uniform periodicity and apodization in the grating structure.  

In a first experiment, we splice together commercially-available FBGs to form a 508\,cm fiber cavity, with a ``round trip'' time of nearly 50\,ns. 
The two FBGs have peak nominal reflectivities of 0.998 and 0.997 at $\lambda_B = 1550$\,nm, with FWHM reflection bandwidths of 0.8\,nm ($100$\,GHz at 1550\,nm). 
Using bright pulses, resonant at the cavity wavelength, we measure a cavity ringdown with $1/e$ lifetime of 12 round trips (see Fig.~\ref{fig:setup}(a, right axis)).
Neglecting other sources of loss, we estimate the loss due to the fiber splice to be 0.17\,dB.
For memory operation the signal pulse ($s$), attenuated to one photon per pulse on average, enters the fiber cavity simultaneously with two \emph{write} control pulses, labelled $p$ and $q$.
After transmitting through the first FBG, the signal is frequency-converted into the reflection band of the FBG via Bragg-scattering four-wave mixing (BSFWM)~\cite{Mckinstrie2005} (see Fig.~\ref{fig:setup}(b-c)).
The converted signal ($t$) is stored in the fiber cavity (Fig.~\ref{fig:setup}(d)) until, after a pre-determined storage time, two \emph{read} control pulses ($p$ and $q$) enters the cavity and drives the reverse BSFWM process (Fig.~\ref{fig:setup}(e)).
The stored photon is converted back to its original frequency,  transmits through the second FBG, and exits the fiber.

The cavity was fabricated in a single-mode fiber (Nufern 1060-XP) which has its zero-dispersion wavelength $\lambda_{ZD} \sim 1435$\,nm. 
For BSFWM in non-birefringent fiber the signal and pump frequencies are on opposite sides about the zero-dispersion frequency such that the signal ($s$) and first control ($p$) are group-velocity matched in the fiber, as are the cavity target mode ($t$) and second control ($q$). 
The frequency splitting between the two control fields sets the splitting between the signal and stored fields, i.e., $\omega_p - \omega_q = \omega_t - \omega_s = 2\pi \times 4.3$\,THz.
For the wavelengths shown in Fig.~\ref{fig:setup}(b), the group velocity walkoff between the two controls is approximately $0.7$\,ps over the 508\,cm fiber cavity, much less than their 10\,ps pulse durations.
This splitting is also large enough to suppress cascaded Bragg scattering of the signal into higher frequency modes~\cite{Christensen2018}.

%%% FIG %%%
\begin{figure*}[ht]
\center{\includegraphics[width=2.\columnwidth]{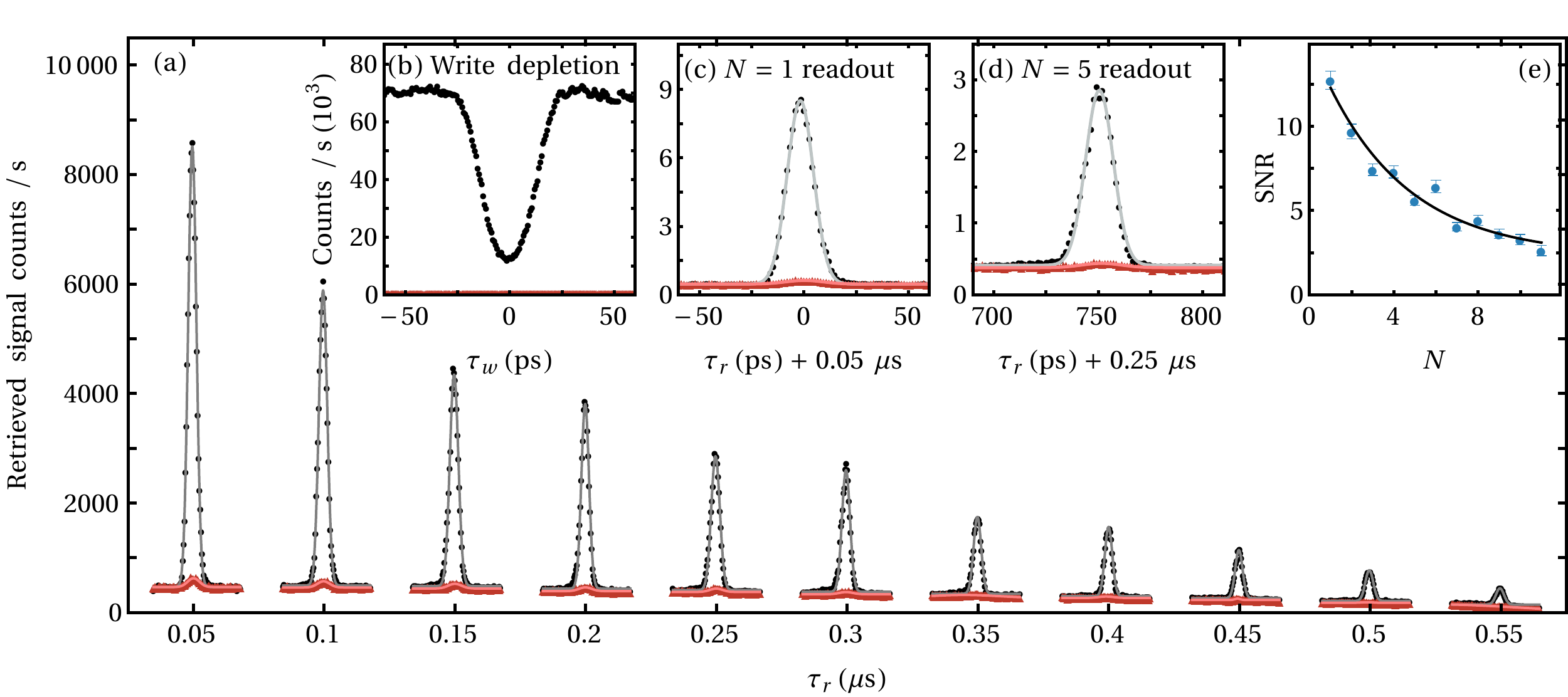}
\caption{ 
Spliced-fiber cavity storage results. 
(a) Retrieved signal (black) and noise (red) photon detections after 1-11 round trips of storage.
Read pulse delay time $\tau_r$ is scanned across a $\sim 100$\,ps range to meet the signal after each 50\,ns round trip. 
After 1 round trip, the total memory efficiency is 11.3\%. 
(b) Write control pulse delay scan. A dip in signal counts at time $\tau_w = 0$\,ns occurs when write pulses are temporally overlapped with the input signal.
(c-d) Zoomed-in read delay scans for (c) $N = 1$, and (d) $N = 5$ round trips, corresponding to 0.05\,$\mu$s and 0.25\,$\mu$s of storage, respectively.
(e) The signal-to-noise ratio (SNR) at the maximum of each read delay scan.
An exponential decay is fit to the data with a $1/e$ decay constant of 4 round trips.
}\label{fig:teraxread}
}\end{figure*}

The experimental setup is shown in Fig.~\ref{fig:setup}(f).  
The primary laser for this experiment outputs 10\,ps duration pulses with center wavelength at 532\,nm and an 80.1\,MHz repetition rate.
The 532\,nm light is split to pump two optical parametric oscillators (OPO), from which the two control fields $p$ and $q$ are derived.
A portion of 532\,nm light also pumps a 20\,mm periodically-poled lithium niobate (PPLN) crystal which, through difference frequency generation with a 800.8\,nm diode laser (\emph{Thorlabs} L808P010), produces bright signal pulses at 1585\,nm.
Signal pulses are filtered through an etalon to a FWHM bandwidth of 34\,GHz and coupled into the fiber cavity with 57\% efficiency. 
For the duration of the experiment, the signal is attenuated to $0.99\pm 0.02$ photons per pulse, as measured at the output of the fiber cavity. 
Control pulses $p$ and $q$, with wavelengths $\lambda_p = 1312$\,nm and $\lambda_q = 1337$\,nm, are combined on a PBS and co-polarized.
Two Pockels cells (PC) are used, in conjunction with crossed-polarizers, to gate \emph{write} and \emph{read} control pulse pairs into the experiment. 
We pick write and read pulses every 2\,$\mu$s, to allow the fiber cavity to completely ring down in between trials. 
Write and read pulses pick up respective delays $\tau_w = 0$\,ns and $\tau_r = N \times 50$\,ns, where $N$ is an integer representing the number of cavity round trips.
The $p$ and $q$ fields are temporally- and spatially-overlapped with the signal on a short-pass dichroic mirror, and coupled into the fiber cavity.
The pulse energies of $p$ and $q$ in the fiber are both 1.7\,nJ. 
Upon exiting the fiber, the control fields are removed by long-pass edge filters. 
The signal is collected into a single-mode fiber, sent through a variable 4$f$-filter (\emph{II-VI} waveshaper) passing a 0.3\,nm-wide band centered at 1585\,nm, and directed to superconducting nanowire signal-photon detectors (SNSPDs).
Detection events with 100\,ps timing resolution are correlated with PC triggers on a coincidence logic unit.

The results for the spliced-fiber cavity memory, are shown in Figure~\ref{fig:teraxread}.
When the input signal ($s$) overlaps in time with the write pulse, we find an 83\% depletion of the input light (see Fig.~\ref{fig:teraxread}(b)). 
The quantum frequency conversion efficiency to the storage wavelength ($\lambda_t = 1550$\,nm) is approximately 40\%; this is confirmed by tuning $\lambda_q$ to convert the signal to a wavelength away from the FBG reflection band.
The remaining depletion is due to cross-phase modulation which broadens the signal to wider than the 0.3\,nm spectral filtering. This is caused by the intense control pulses which have a combined pulse energy of 3.4\,nJ.

Figure~\ref{fig:teraxread}(a) shows detection rates of the retrieved signal for up to $N=11$ round trips (0.55\,$\mu$s) of storage. 
Noise generated by the control pulses, when no input signal is sent, is plotted alongside the readout data.
Reading out the signal after one round trip (Fig.~\ref{fig:teraxread}(c) shows a zoom-in), we observe a strong peak when the read pulse delay matches the circulating stored signal pulse, converting it back to 1585\,nm. 
We find a total memory efficiency of $(11.3 \pm 0.1)\%$  after one round trip, defined as the retrieved signal rate ($N=1$ peak) divided by the input signal rate (baseline in Fig.~\ref{fig:teraxread}(b)).

%%% FIG %%%
\begin{figure}
\center{\includegraphics[width=1\columnwidth]{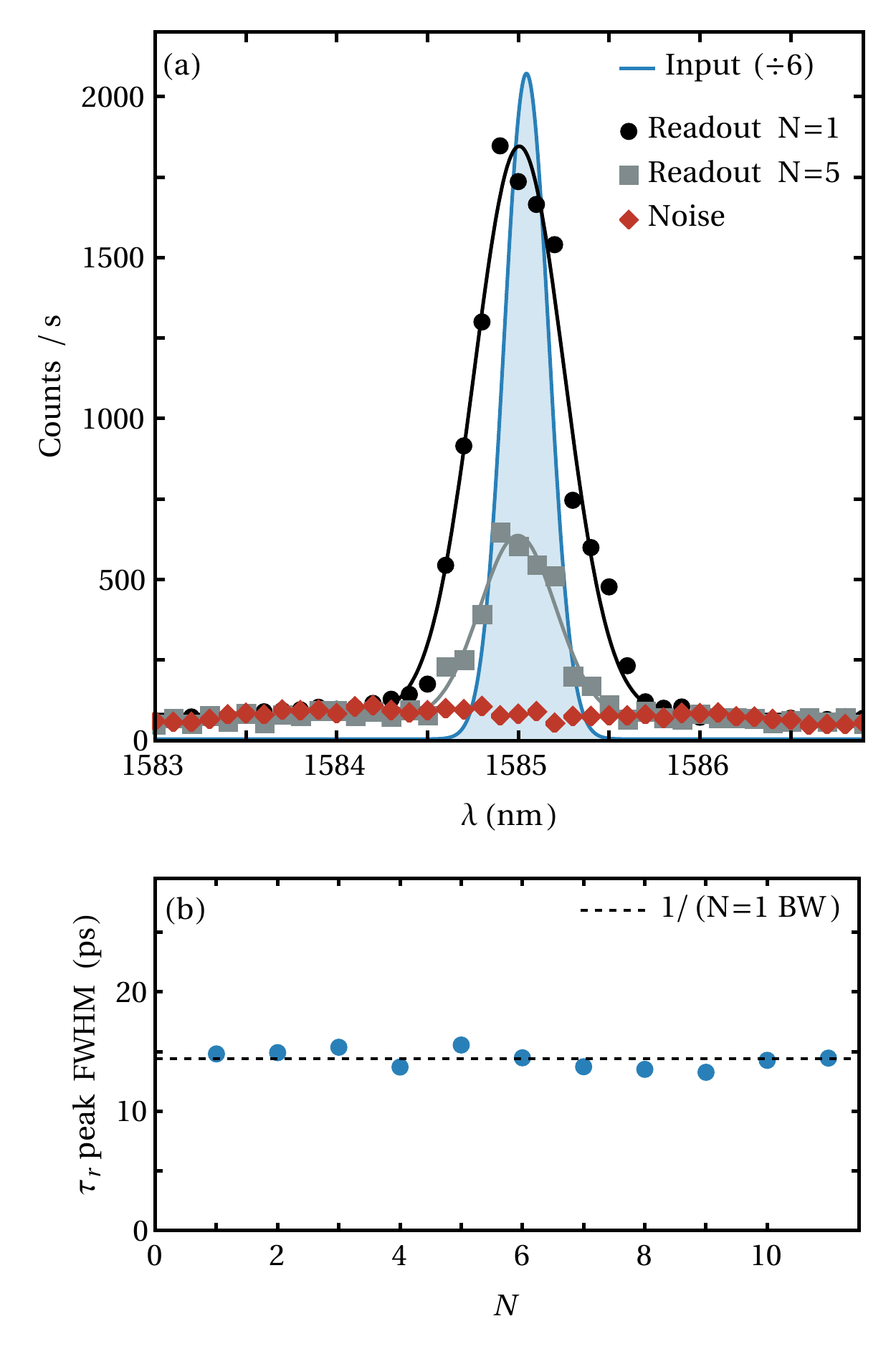}
\caption{ (a) Spliced-fiber cavity - spectra of retrieved signal after 1 and 5 round trips of storage, compared against the input (divided by 6 for clarity) and noise. The spectral fidelities with the input are $\mathcal{F}_\lambda=0.896$ and $\mathcal{F}_\lambda=0.895$ for $N=1$ and $N=5$, respectively.
(b) FWHM of peaks in $\tau_r$ delay scan for each round trip. Dashed line is the transform-limit for the measured $N=1$ readout spectrum. 
} \label{fig:teraxspectra}
}\end{figure}

%%% FIG %%%
\begin{figure}
\center{\includegraphics[width=1.\columnwidth]{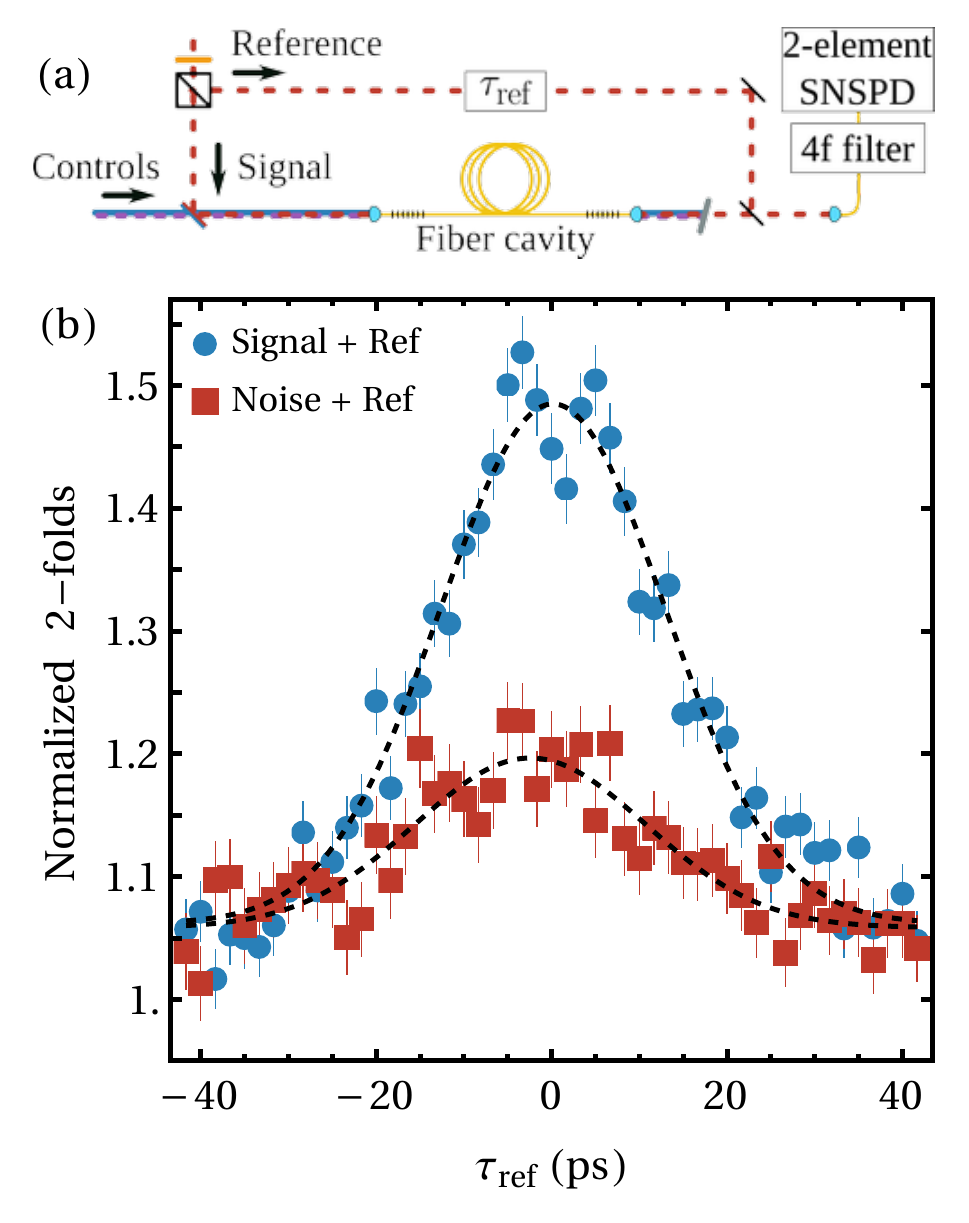}
\caption{ (a) Setup to observe interference between a delayed reference signal, and a signal retrieved from the memory after 1 round trip. They are combined on a PBS, diagonally polarized and incident on a two-element SNSPD.
(b) Spliced-fiber cavity - normalized two-fold detections on the two-element SNSPDs shows interference between readout ($N=1$) with the delayed reference signal.  Fit of main peak (blue) shows a 42\% increase in normalized two-fold detections when reference and readout paths are overlapped. This is consistent with the maximal interference visibility scaled by the spectral fidelity.
} \label{fig:teraxHOM}
}\end{figure}

At each round trip, the read control pulse delay ($\tau_r$) is scanned, in 1.33\,ps steps, across a 100\,ps range to match the delay of the stored signal pulse circulating in the fiber cavity.
We retrieve the signal for longer storage times by shifting the timing of the control pulse picking in 50\,ns increments.
We find that the memory efficiency decays faster than one would expect from the 12-round trip cavity lifetime as measured in Fig.~\ref{fig:setup}(b, right axis).
This is due to a discrepancy between the fiber cavity round trip time and the time between control pulses from the laser.
The measured cavity round trip time is 49.69\,ns, and so the stored signal arrives at the same point in the fiber cavity approximately 200\,ps earlier than the read pulse.
Therefore, for each subsequent round trip $\tau_r$ is adjusted by 187\,ps to compensate. 
This changes the read beam path by over 50\,cm between $N=1$ and $N=11$, and, by consequence, the coupled read pulse energy into the fiber cavity is reduced by a factor of 3, reducing the overall memory efficiency.
In Section~\ref{sec:monolithic}, we overcome this hurdle using a monolithic fiber cavity which can be mechanically stretched into resonance with the primary oscillator. 

The main source of noise in this experiment is from Raman scattering of read control pulses in the doped-silica fiber. 
The silica Raman spectrum is well-known and a small, but non-neglible, amount of noise is present at 31\,THz and 35\,THz, the frequency differences between the nearest pump field ($\lambda_q = 1337$\,nm) and the cavity and signal fields, respectively.
Raman noise generated in the cavity in the write step  can be subsequently mapped out of the cavity in the read step.
This results in a time dependent noise signal which gives a peak in the read pulse delay scan, which is observable for up to $N = 6$ round trips. 
Despite this noise, we observe a high signal-to-noise ratio (SNR), which remains above unity for $N=11$ round trips of storage (see Fig.~\ref{fig:teraxread}(e)).

Next, we measure the spectrum of the retrieved signal by scanning the $4f$-filter in 0.1\,nm steps.
Figure~\ref{fig:teraxspectra}(a) plots the measured spectra of the input signal, the retrieved signal after $N=1$ and $N=5$ round trips of storage, as well as the noise.
We find that the retrieved signal spectrum at $N=1$ is 2 times broader than the input spectrum. Similarly, the retrieved signal spectrum at $N=5$ is 1.7 times broader than the input. 
We expect this broadening is due to cross-phase modulation of the input light in the frequency conversion process. 
A crucial metric for any quantum memory is the fidelity of the retrieved signal with the input. 
Computing the spectral fidelity as $\mathcal{F}_\lambda = \int d\lambda \sqrt{I_\text{in}(\lambda) I_\text{out}(\lambda) } / \left  [ \int d\lambda I_\text{in}(\lambda) \int d\lambda I_\text{out}(\lambda)  \right ]^{1/2}$, we find $\mathcal{F}_\lambda = 0.896$ between the input and noise-subtracted readout signals at $N=1$, and  $\mathcal{F}_\lambda = 0.895$ at $N=5$.  
We note that the temporal profiles of the retrieved signal pulses remain consistent for all measured storage times. 
The fitted signal peak widths, when scanning $\tau_r$, are plotted in Fig.~\ref{fig:teraxspectra}(b).  
The dashed line through the data is the inverted bandwidth of the $N=1$ retrieved signal spectrum, showing that retrieved signals remain near the transform limit. 

To assess the performance of this memory for potential photon source multiplexing, we perform a Hong-Ou-Mandel~\cite{Hong1987} interference experiment between a photon retrieved from the memory ($N=1$) and a reference photon delayed outside the cavity for $\sim 50$\,ns.
The reference and retrieved signal are combined on a PBS, diagonally-polarized and, after collection into SM fiber and spectral filtering, are sent to a two-element SNSPD. 
By measuring coincident detections from the two elements, which is equivalent to a beamsplitter followed by two detectors, we observe photon bunching between the reference and retrieved signal when their path lengths are matched.
This is equivalent to Hong-Ou-Mandel interference between two coherent states, where only one of the beamsplitter output ports is measured. 
Two indistinguishable single photons meeting at a 50:50 beamsplitter bunch so that, while one expects a decrease in coincident detections between the output ports, there is an increase in two-photon events in each beamsplitter output port.
This increase can be observed with a Hanbury Brown-Twiss setup~\cite{HBT1956}, or, in our case, a two-element SNSPD.  
In Figure~\ref{fig:teraxHOM} we plot the normalized two-fold detections, $N_{1,2} / (N_1 N_2)$, as a function of the reference delay, $\tau_\text{ref}$, where $N_i$ is the number of detector clicks from each of the two elements.
We see a peak in normalized two-folds when the reference and retrieved signal overlap in time. 
A Gaussian fit to the data gives an amplitude of 0.423, where the maximum allowed value in classical optics is 0.5~\cite{MandelWolf}. 
Our measured peak amplitude is 84.6\% of the maximum which, when considering that Raman noise is present in this scan, is in good agreement with the spectral fidelity of $\mathcal{F}_\lambda = 0.896$.
We expect this to increase with improved temporal matching of the signal to the pumps.

In this section, we have shown that the spliced-fiber cavity memory can operate up to $N=11$ round trips with total memory efficiency of 12\% and spectral fidelity of 0.896 after 1 round trip.  
Two issues arose in this experiment which both trace back to the fiber splice.
The first is that the cavity ringdown was much shorter than expected given the nominal FBG reflectivities (0.998 and 0.997).
We estimated the splice loss to be 0.17dB, meaning that without a splice in this cavity one could expect cavity ringdowns of well over 100 round trips. 
Second, the cavity round trip time did not match the time between subsequent control pulses.
We accounted for this discrepancy by changing the optical path delay of the read pulse, but this is not sufficient for the proposed applications of this memory, i.e., the synchronization of photon sources.
Coupled read pulse energy, and consequently memory efficiency, also dropped as a result.
While the fiber cavity can be mechanically stretched to increase the round trip time, this would likely damage the splice.
We address both these issues in the following section.

%%% FIG %%%
\begin{figure*}[ht]
\center{\includegraphics[width=2.\columnwidth]{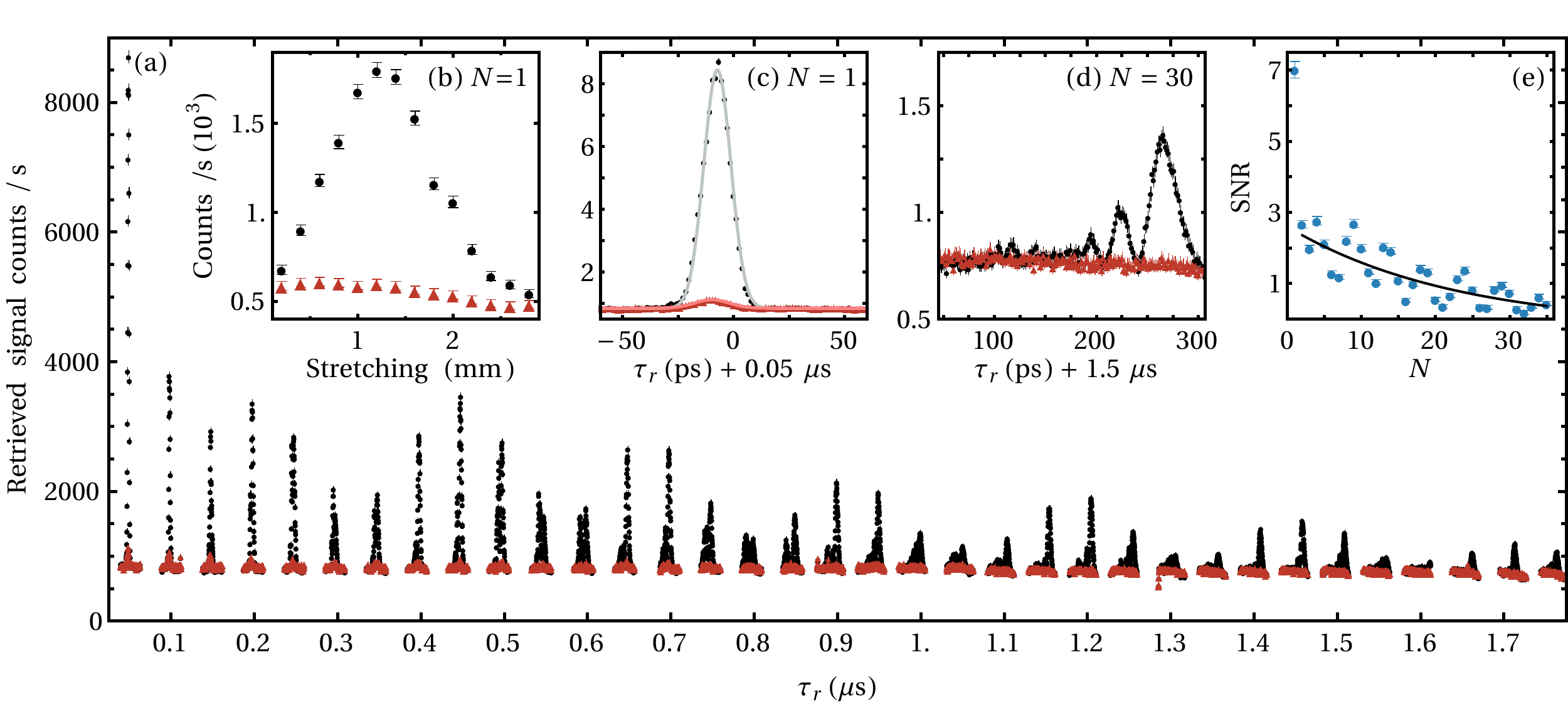}}
\caption{ Monolithic cavity storage results.  (a) Retrieved signal (black) and noise (red) photon detections after $N=1-35$ round trips of storage in the monolithic fiber cavity.
Read pulse delay time $\tau_r$ is scanned across a 100-300\,ps range to meet the signal after each 50\,ns round trip. 
After one round trip, the memory is $12.2 \%$ efficient. 
(b) Mechanical stretching of the cavity across a 3\,mm range shows while the memory is operating show a peak when it is brought into resonance with the control laser repetition rate. 
(c-d) Zoomed in read delay scans for (c) $N = 1$ and (d) $N = 30$ round trips, corresponding to 0.05\,$\mu$s and 1.5\,$\mu$s of storage, respectively. 
(e) The signal-to-noise ratio (SNR) at the maximum of each round trip scan. 
An exponential decay is fit to the data with a $1/e$ decay constant of 22 round trips.
}\label{fig:cyril_read}
\end{figure*}

\section{ Monolithic fiber cavity }
\label{sec:monolithic}
In a second experiment, we construct a monolithic cavity by writing two 3.5\,mm-long FBGs at either end of a single-mode fiber (1060-XP). 
The type-I FBGs were inscribed using femtosecond pulses at 800\,nm from a 1\,KHz Ti:sapphire amplifier~\cite{Mihailov2003, Mihailov2012}.
By chirping the grating, i.e., sweeping the grating period $\Lambda$ across a few-nm range over the length of the grating, the reflection bandwidth can be increased.  
With a chirped FBG at one end of the cavity and an anti-chirped FBG at the other, a trapped pulse acquires opposite signs, ideally in equal amounts, of dispersion from each FBG reflection. 
In the monolithic cavity the FBGs are chirped such that the grating period $\Lambda$ is swept by 5\,nm across the length of the FBGs.
The FBGs have a 4\,nm-wide reflection band centered on 1546.5\,nm, with reflectivities of 0.990 at 1547.2\,nm.
We measured a $1/e$ cavity ringdown of 49 round trips, or 2.45\,$\mu$s, much longer than in the spliced-fiber cavity. 
However, these longer ringdowns occur near the red edge of the FBG reflection band, whereas the blue side of the reflection band experiences loss due to coupling to fiber cladding modes~\cite{Mizrahi1993}. 
In this experiment, we also clamp the fiber cavity near each FBG and apply tension across the cavity by scrolling a translation stage.
By stretching the fiber over a 3\,mm range, we bring the memory cavity into resonance with the primary laser cavity.
This is crucial for the synchronization of photon sources.

We repeat the same storage and retrieval experiment in this fiber cavity, tuning the control wavelengths to $\lambda_p = 1310$\,nm and $\lambda_q = 1337$\,nm to account for the different FBG reflection and dispersion properties of the fiber.
In the write process we measure a 75\% depletion of the input signal which, similarly to the spliced-fiber case, is partially due to cross-phase modulation broadening the signal to wider than the 0.3\,nm spectral filtering. 
Figure~\ref{fig:cyril_read}(a) shows rates of retrieved signal photons from the memory for up to $N=35$ round trips (1.75\,$\mu$s) of storage. 
This is over 3 times longer that the storage lifetime observed in the spliced-fiber cavity. 
Figure~\ref{fig:cyril_read}(b) shows a peak in retrieved photon rates when mechanically stretching the cavity across a few-mm range, and leaving the control pulse timing unchanged. 
This shows that we are able to bring the cavity into resonance with the primary laser in the experiment so that control pulses can address different integer round trips of storage with electronic control of pulse picking, rather than by also changing the optical delay of the read pulses.  
This also implies that the read efficiency from the memory remains higher for longer storage times. 
After 1 round trip (see Fig.~\ref{fig:cyril_read}(c)), we measure a $(12.7 \pm 0.2)\%$ total memory efficiency, and an SNR of $7.0 \pm 0.2$. 

%%% FIG %%%
\begin{figure}
\center{\includegraphics[width=1.\columnwidth]{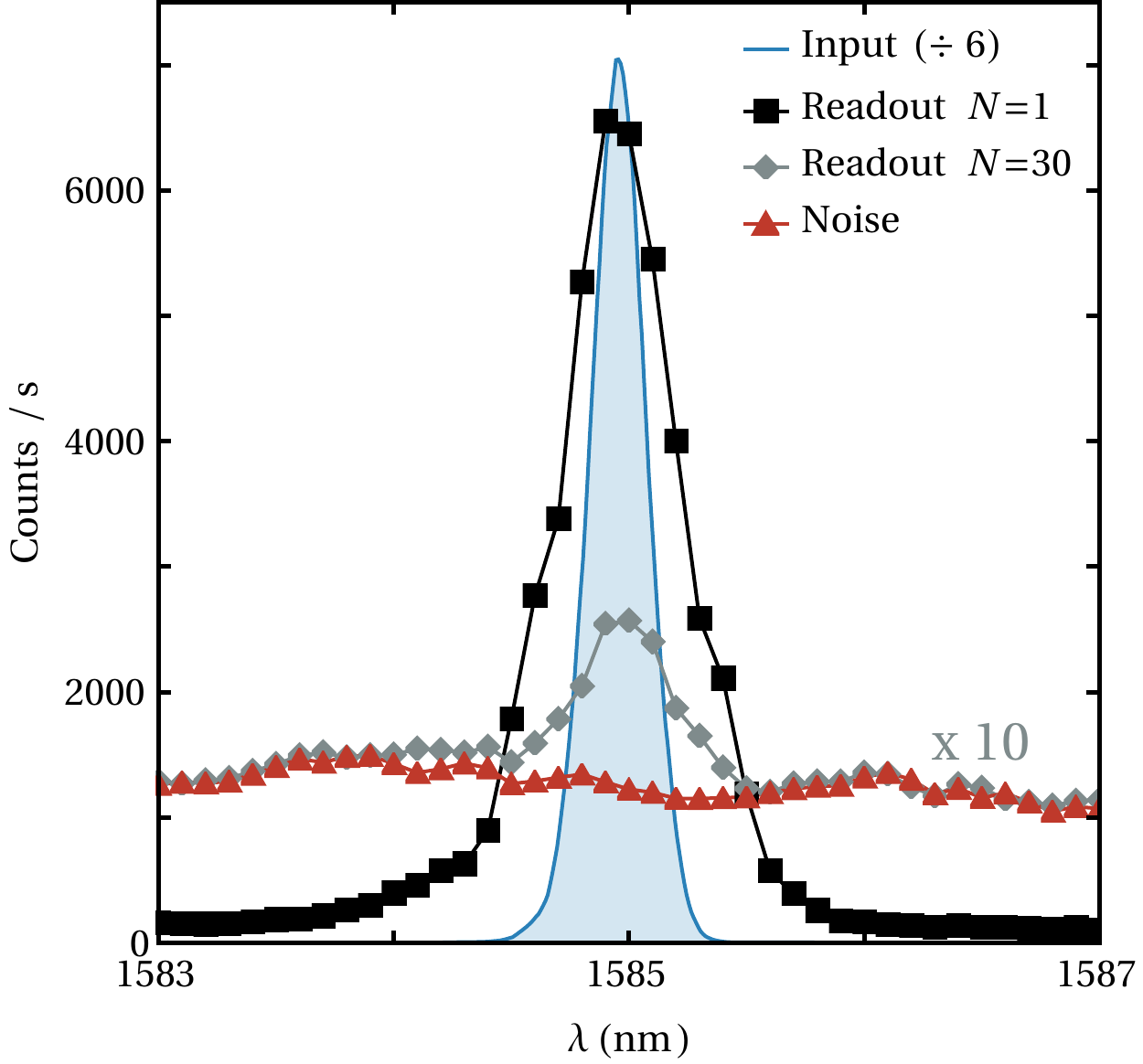}}
\caption{(a) Monolithic cavity - spectra of retrieved signal after 1 and 30 round trips of storage, compared against the input (divided by 6 for clarity) and noise. The spectral fidelities with the input are $\mathcal{F}_\lambda=0.875$ and $\mathcal{F}_\lambda=0.868$ for $N=1$ and $N=30$, respectively.
Note that the $N=30$ readout and noise data are scaled by a factor of 10 for clarity. 
}\label{fig:cyril_spec}
\end{figure}

%%% FIG %%%
\begin{figure}
\center{\includegraphics[width=1\columnwidth]{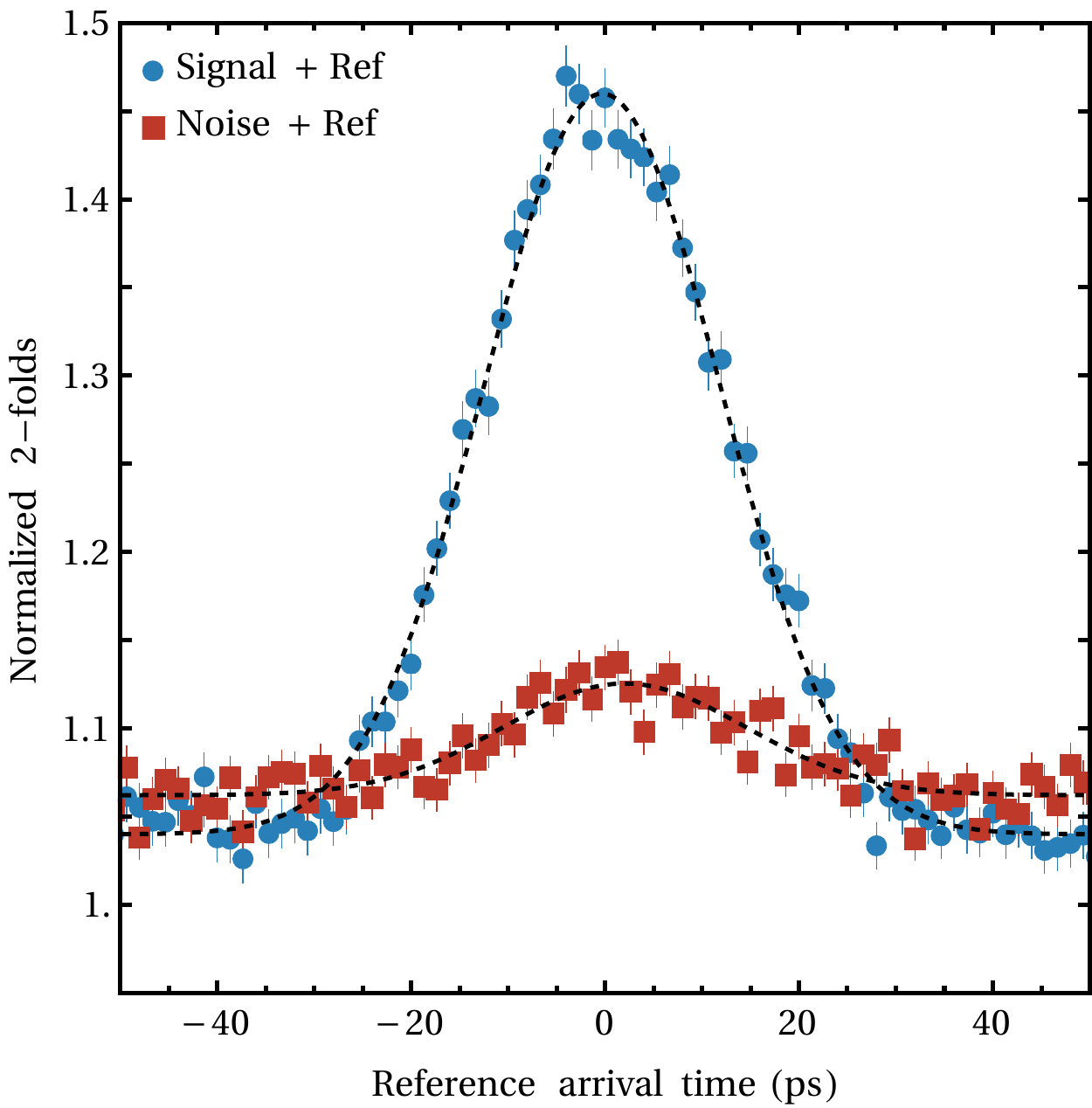}}
\caption{Monolithic cavity  - interference between $N=1$ retrieved signal with a delayed reference pulse. Fit of the signal peak (blue) shows a 40\% increase in normalized two-fold detections when reference and signal paths are overlapped.}
\label{fig:cyril_HOM}
\end{figure}

We find that for many $N$ in the readout scan in Fig.~\ref{fig:cyril_read}(a), shown in detail for $N=30$ in Fig.~\ref{fig:cyril_read}(d), the retrieved photon peaks significantly broaden and acquire additional features.
In particular, the profile for $N=30$ is indicative of third-order dispersion acting on a pulse, which we attribute to the FBGs. 
While the FBGs for the monolithic cavity were designed to mitigate third-order dispersion, it is clear that it is being applied to stored photons. 
Figure~\ref{fig:cyril_read}(e) plots the SNR over the 35 round trips of storage, showing an initial drop from the value of 7 at $N=1$, followed by an exponential decay with a $1/e$ constant of 22 round trips. 
The exponential decay also shows an oscillation with a period of 4-5 round trips, which can also be seen in the readout peaks in Fig.~\ref{fig:cyril_read}(a). 
This is mainly due to the polarization of the stored signal rotating as it circulates in the cavity. 
Since the read control pulses are co-polarized with the write controls, as well as the input signal, the readout frequency conversion occurs with oscillating efficiency as the stored signal changes polarization. 
We did not observe this oscillation in the spliced-fiber cavity, and it may be caused by inducing birefringence when stretching the fiber cavity. 
This makes it clear that next-generation devices should be fabricated in polarization-maintaining fiber~\cite{Bustard2022}, which has the added feature of more phase matching options in the frequency conversion process. 

We measure the spectra of retrieved photons from the monolithic cavity after $N=1$ and $N=30$ round trips, plotted in Fig.~\ref{fig:cyril_spec}.
We find, as for the spliced-fiber cavity, that the retrieved signal spectra are broader than the input spectrum, and we attribute this broadening to cross-phase modulation occurring in the four-wave mixing processes for the read and write operations. 
The spectral fidelities of the input signal the retrieved signal after $N=1$ and $N=30$ round trips are 0.875 and 0.868, respectively. 
Even though the temporal profile of the stored light is being distorted over longer storage times, the spectral fidelity of the memory remains consistent for longer storage times. 
Lastly, we again interfere the retrieved $N=1$ signal with a delay reference signal after the memory. 
Figure~\ref{fig:cyril_HOM} shows the peak in normalized two-fold detections when the reference pulse is scanned through the retrieved signal.
A Gaussian fit to the data gives an amplitude of 0.42, or 84\% of the maximum allowed value, in close agreement with the measured spectral fidelity when considering the additive noise. 

\section { Discussion \& Outlook}

In this work we have developed a broadband quantum memory for telecom light based on a fiber cavity. 
We have demonstrated the storage and retrieval of telecom single-photon-level pulses for microsecond timescales in two separate devices. 
In the first, spliced-fiber cavity, we observed low dispersion for the stored photon, while in the second, monolithic cavity, we observed longer lifetimes and the ability to stretch the cavity into resonance with the primary laser. 
The memories operated with 11.3\% and 12.7\% efficiencies, respectively, after one round trip. 
This efficiency can increase with more available pump pulse energy.
We expect that the combination of low-dispersion FBGs with a monolithic design will result in a long-lived, high-fidelity memory.
Next-generation devices will also focus on mitigating noise generated by the control pulses during the read and write operations.
We expect that Raman noise can be greatly reduced by swapping the wavelengths of controls and signals in this experiment.
This will precipitate the storage of heralded single photons with high signal-to-noise, making this memory immediately useful for local quantum networking applications.

While microsecond lifetimes are insufficient for long-distance quantum networking, we expect this memory to be useful for the synchronization of photon sources. 
In particular, we expect fiber-based cavity memories to closely follow their free-space counterparts in providing rate enhancement in spontaneous photon generation. 
Rate enhancements of 4- and 6-photon GHZ states~\cite{Meyer-Scott2022} were achieved in a free-space cavity with a $1/e$ lifetime of 10 round trips. 
Similarly, 30-fold enhancements by time-multiplexing a heralded single-photon source were demonstrated in a free-space cavity with a $1/e$ lifetime of 83 round trips~\cite{Kaneda2017, Kaneda2019}.  
The spliced- and monolithic fiber cavities demonstrated in this experiment had $1/e$ lifetimes of 12 and 49 round trips, respectively.
We expect that for longer storage times that the fiber-based cavities will also exhibit desirable spatial modes, facilitating high-visibility interference with independently-generated photons upon retrieval from the memory.

\begin{acknowledgments}
We are grateful for discussions with Khabat Heshami, Aaron Goldberg, Fr\'{e}d\'{e}ric Bouchard, Kate Fenwick, Guillaume Thekkadath and Yingwen Zhang. We also thank Denis Guay, Rune Lausten, Rob Walker and Doug Moffatt for technical assistance.
\end{acknowledgments}

\bibliography{FBG.bib}
%\bibliography{_PRXsub_20230322}
\end{document}